# Linearizing nonlinear optics

Bruno E. Schmidt[1,2], Philippe Lassonde[2], Guilmot Ernotte[2], Matteo Clerici[3], Roberto Morandotti[2], Heide Ibrahim[2] and François Légaré[2]

[1]*Few-cycle Inc., 2890 Rue de Beaurivage, Montreal, H1L 5W5, Qc, Canada*
[2]*INRS-EMT, 1650 Blvd. Lionel Boulet, Varennes,*
[3]*University of Glasgow, School of Engineering, G12 8QQ, Glasgow, England*
schmidt@few-cycle.com
legare@emt.inrs.ca
17.03.2016

In the framework of linear optics, light fields do not interact with each other. Yet, when their field amplitude becomes comparable to the electron binding energies of matter, the nonlinear motion of these electrons emits new dipole radiation whose amplitude, frequency and phase differ from the incoming fields. Such high fields are typically achieved with ultra-short, femtosecond (1fs = $10^{-15}$ sec.) laser pulses containing very broad frequency spectra. Here, the matter not only couples incoming and outgoing fields but also causes different spectral components to interact and mix through a convolution process.

In this contribution, we describe how frequency domain nonlinear optics overcomes the shortcomings arising from this convolution in conventional time domain nonlinear optics[1]. We generate light fields with previously inaccessible properties because the uncontrolled coupling of amplitudes and phases is turned off. For example, arbitrary phase functions are transferred linearly to the second harmonic frequency while maintaining the exact shape of the input power spectrum squared.

This nonlinear control over output amplitudes and phases opens up new avenues for applications based on manipulation of coherent light fields. One could investigate *c.f.* the effect of tailored nonlinear perturbations on the evolution of discrete eigenmodes in Anderson localization[2]. Our approach might also open a new chapter for controlling electronic and vibrational couplings in 2D-spectroscopy[3] by the geometrical optical arrangement.

Many engineering and computational physics problems are routinely solved by addressing them in Fourier domain. A paradigmatic example is the differential problem posed by the time dependent Schrödinger equation, a problem of first order in time and second order in space[4]. Applying a spatial Fourier transformation (FT) leads to an ordinary differential equation – a significant reduction of complexity.

We have adapted this approach experimentally by performing nonlinear optics in the frequency domain rather than in the time domain. We show that it is then possible to tailor the spectral properties of the nonlinearly generated field with unprecedented control. We introduce the term *Frequency domain Nonlinear Optics* (FNO) to identify this approach. A general framework with an intuitive model to predict its phenomenology is provided. For the sake of clarity, we restrict the discussion to second harmonic generation (SHG)[5], the first light induced optical nonlinearity observed in 1961 by Franken and co-workers. However, our concept applies to other nonlinear effects as was recently demonstrated for broadband parametric amplification of few-cycle laser pulses[6]. There, it challenged a longstanding paradigm stating that in laser media an increased laser gain typically compromises the amplified bandwidth. Now, it is possible to increase the gain by increasing the bandwidth.

The basic physics underlying SHG can be presented in simple terms: Upon the excitation of an optical field $\mathcal{E}(t)$[1], *e.g.* a pulse at carrier frequency $\omega_0$, a nonlinear polarisation wave is excited in the medium[7]. For a non-centrosymmetric medium, the nonlinear polarisation is proportional to the product of the input electric fields in the time domain: $P^{(2)}(t) = \epsilon_0 \chi^{(2)} \mathcal{E}^2(t)$. The fast oscillating component of this field acts as a source for electromagnetic radiation at twice the input frequency:

$$E_T^{SH}(t) \propto E_T(t) \cdot E_T(t). \tag{1}$$

The subscript $T$ identifies this type of interaction as the conventional *Time domain Nonlinear Optics* (TNO). The spectral representation of Eq. (1) can be obtained via a mathematical FT:

$$E_T(2\omega) \propto E_T(\omega) * E_T(\omega). \tag{2}$$

The multiplication of E-fields in time representation changes to a convolution operation (*). Eq. (2) denotes the spectral representation of TNO where $E_T(\omega)$ contains the entire pulse spectrum, as shown in Fig. 1(a). The convolution operation mixes all spectral components of the input pulse, such that the SH spectral amplitude and phase are dependent on the spectral amplitude and phase of all frequency components of the input pulse. Therefore, the SH output spectrum in Fig. 1(b) can differ significantly from the input.

The key aspect of FNO is to achieve a very narrow bandwidth for the nonlinear light-matter interaction which is achieved experimentally by means of a *4-f* optical setup[8,9] as described in Figure *1*(c). The experimental setup visualizes the basic equation of FNO:

$$E_F(2\omega) \propto \int_{-\infty}^{\infty} d\omega' \left(E_F(\omega) \cdot \delta(\omega - \omega')\right) * \left(E_F(\omega) \cdot \delta(\omega - \omega')\right). \tag{3}$$

---
[1] We employ the complex notation for the electric field: $\mathcal{E}(t) = Re[E(t)]$

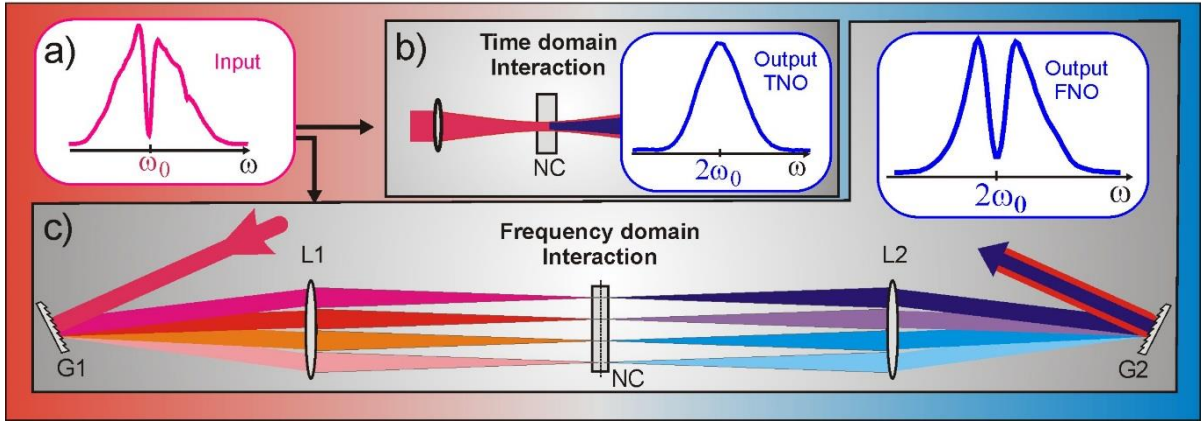

Figure 1: An input spectrum corresponding to transform limited 33 fs pulses at 800 nm wavelength exhibiting a spectral hole (a) is frequency doubled via Time domain Nonlinear Optics (TNO) in (b) and Frequency domain Nonlinear Optics (FNO) in (c). In both cases, we use a 150 μm thin BBO as the nonlinear crystal (NC) to minimize effects of phase mismatch. In TNO (b), all frequencies interact simultaneously in a single focus and the experimental SH spectrum (blue curve) shows a smooth function without central hole, a result of the convolution described in Eq. (2). In FNO (c), the frequencies of an incident pulse disperse after the first grating G1 and their different propagation angles become parallel after the first lens L1. The lens again focuses each plane wave beamlet into a small focal spot next to its neighbouring frequency. All frequency components side by side form the frequency plane at one focal length behind the lens. This setup performs a continuous optical Fourier transformation from the time domain to the frequency domain. A symmetric setup subsequent to the frequency plane performs a second Fourier transformation back to the time domain. The experimentally obtained time domain SH spectrum (blue curve) exhibits the same shape as the input spectrum.

The optical FT provides a reduced frequency content in each focal spot of the frequency plane (FP) which is expressed by the term $E_F(\omega) \cdot \delta(\omega - \omega')$, where $\delta(\omega)$ denotes the Dirac delta function. The subscript $F$ refers to the physical interaction in the frequency domain. The integration over all frequency components ($\int d\omega'$) corresponds to the coherent recombination of all frequencies at the second grating G2 to form collimated output beams of the fundamental and SH (see methods section). Note that the convolution representing the nonlinear interaction is still present in Eq. (3). It applies, however, only to isolated, single frequencies. Thus, the mutual coupling between frequencies is turned off. Evaluating Eq. (3) leads to (see *SI*):

$$E_F(2\omega) \propto E_F(\omega) \cdot E_F(\omega) = S^2(\omega)e^{i2\phi(\omega)}, \qquad (4)$$

where the spectral amplitude $S(\omega) \equiv |E(\omega)|$ and spectral phase $\phi(\omega) \equiv Arg[E(\omega)]$ substantially differ from Eq. (2). The convolution of E-fields has changed to a multiplication. This crucial difference is the foundation of FNO.

*Power spectrum*

We start the discussion of the far-reaching consequences by comparing the experimental power spectra obtained via Time domain SHG (TSH) and Frequency domain SHG (FSH) in Fig. 1. We first generate a transform limited (TL: flat spectral phase) input pulse with a hole in the centre of its power spectrum (a). It is frequency doubled in a thin nonlinear crystal (BBO). The resulting power spectrum of TSH shown in (b) exhibits a smooth shape, a consequence of the mutual nonlinear mixing of all frequencies described by Eq. (2).

Exploiting the setup drawn in Fig. 1(c) we generate an FSH pulse. For the same input pulse and crystal as in the time domain experiment, we obtain a completely different result: in the SH spectrum the central hole is preserved. Unlike in TSH, arbitrary amplitude transfers from the fundamental to the SH pulse are possible in FSH. It is a direct consequence of Eq. (4), which relates the SH power spectrum with the input power spectrum via a phase independent relation that maintains a direct correspondence between input and output frequencies: $I[2\omega] = S^4[\omega]$. We also note that the generated SH output after the *4f* setup exhibits good spatial properties, witnessed by the absence of spatial chirp and a perfectly symmetric far field profile (see *SI*).

*Spectral phase*

Performing TSH of a linearly chirped pulse, *e.g.* a Gaussian pulse with quadratic spectral phase, leads to a linearly chirped Gaussian pulse with a quadratic spectral phase (see SI). Its power spectrum, however, depends on the input chirp. The situation changes drastically when considering higher order phase functions. In that case, the phase resulting from TSH is distorted. Therefore, the standard time domain approach fails to provide the means for arbitrary spectral phase transfer between the input and nonlinear output. However, we prove that FNO complies with this requirement because of the linear input-output relation of the spectra in Eq. (4). Fig. 2 provides an intuitive picture of the difference between TNO and FNO when comparing the pulse spectrograms of a pulse exhibiting third order dispersion (TOD). It shows numerical (a-c) and experimental (d-f) spectrograms that display the instantaneous pulse spectra for each point in time as a 2-d map.

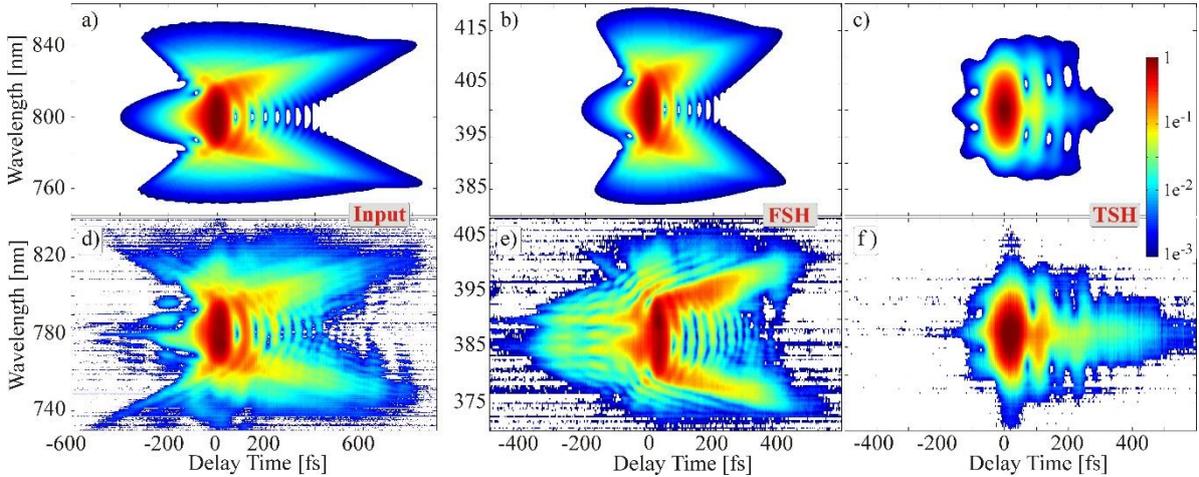

Figure 2: Theoretical (a-c) and experimental (d-f) spectrograms for TSH and FSH for pure phase shaping. Experimentally, the pulses are characterized via transient grating-frequency resolved optical gating[10]. The spectrogram of a TL pulse has the shape of a symmetric balloon without distinct features. A pure TOD of 150000 fs$^3$ was added to a 33 fs TL fundamental pulse in (a). In the experiment (d), additional higher order distortions are present. The typical shape of a butterfly with open wings illustrates how both spectral side bands are delayed with respect to the central frequency (a & d). This butterfly shape is nicely preserved in FSH (b, e) while being completely different in TSH (c, f). For the latter, the butterfly closes its wings because at the trailing edge of the pulse both spectral side bands of the fundamental are annihilated to generate SH photons only around the central frequency $2\omega_0$. Note the logarithmic color bar where green denotes the ~5% level.

The time domain approach clearly results in a completely different spectrogram (c) compared to the input (a), whereas with FSH, the input (a) and output (b) are virtually the same. The calculations are in very good agreement with experiment (d-e). We can extend this qualitative comparison to a quantitative analysis by retrieving the spectral phase from the measured spectrograms[10]. Figure 3 shows the retrieved fundamental (SH) phase in red (blue & cyan). The FSH phase (blue curve) resembles the shape of the fundamental, thus the arbitrary phase is transferred linearly. However, when comparing the two ordinates, the phase is doubled (to all orders) as predicted by Eq. (4). In contrast, the cyan curve from TSH exhibits no linear correspondence with the fundamental phase.

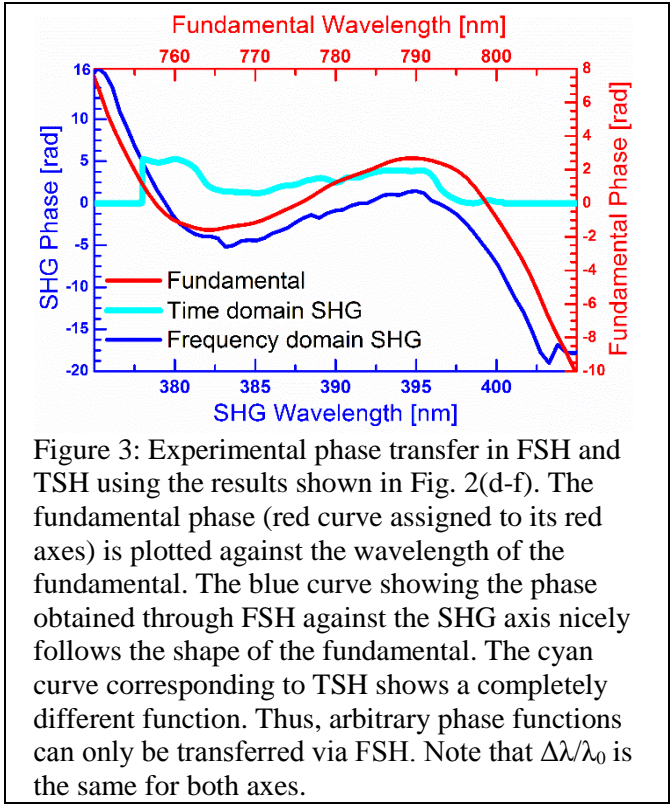

Figure 3: Experimental phase transfer in FSH and TSH using the results shown in Fig. 2(d-f). The fundamental phase (red curve assigned to its red axes) is plotted against the wavelength of the fundamental. The blue curve showing the phase obtained through FSH against the SHG axis nicely follows the shape of the fundamental. The cyan curve corresponding to TSH shows a completely different function. Thus, arbitrary phase functions can only be transferred via FSH. Note that $\Delta\lambda/\lambda_0$ is the same for both axes.

*Phase and amplitude shaping*

We now demonstrate how input pulses with both amplitude and phase variations give different outputs for the two schemes. As mentioned before, even for a Gaussian input pulse with pure linear chirp, the power spectrum obtained through TSH depends on the input spectral phase, i.e. a chirped input generates less SH. We will now discuss how this dependence affects the SH process, and how the amplitude and phase are decoupled by performing the nonlinear interaction in the frequency domain. To this end, and for the sake of clarity, instead of a Gaussian spectrum we consider a spectrally shaped pulse featuring a distinct hole in the spectrum, shown in Fig. 4 (a) as the shaded yellow curve. Its SH is first generated in the time domain, and the resulting power spectrum is recorded for varying linear chirp from $-17600$ fs$^2$ to $17600$ fs$^2$. This stretches a TL pulse from 33 fs to 2.7 ps. The results in Fig. 4 (b) show significant changes of the SH power spectrum as a function of chirp. Again, for a TL pulse (flat phase), the central hole is completely absent (violet curve in Fig. 4 (a)). The central hole only develops as the pulse gets temporally stretched (light grey curve: input stretched to 2 ps). In that case, different frequencies arrive at different times and interferences are suppressed. Furthermore, the SHG efficiency changes significantly, a consequence of the chirp-dependent pulse intensity.

The situation is radically different in the FSH result of (c). The power spectrum obtained for the same input conditions remains the same (black curve in Fig. 4 (a)) irrespective of the input phase. Eq. (4), however, predicts a perfect match of the yellow curve in (a) with the black one since: $I(2\omega) \propto I^2(\omega)$. Yet, the observed discrepancy is explainable. It arises from the limited phase matching bandwidth provided by the birefringence of the 150 µm thick BBO (see SI). Accounting for this macroscopic effect leads to the prediction plotted as the orange curve. The clear one-to-one match for FSH with the phase matched fundamental spectrum ultimately proves the linearization of the nonlinear process.

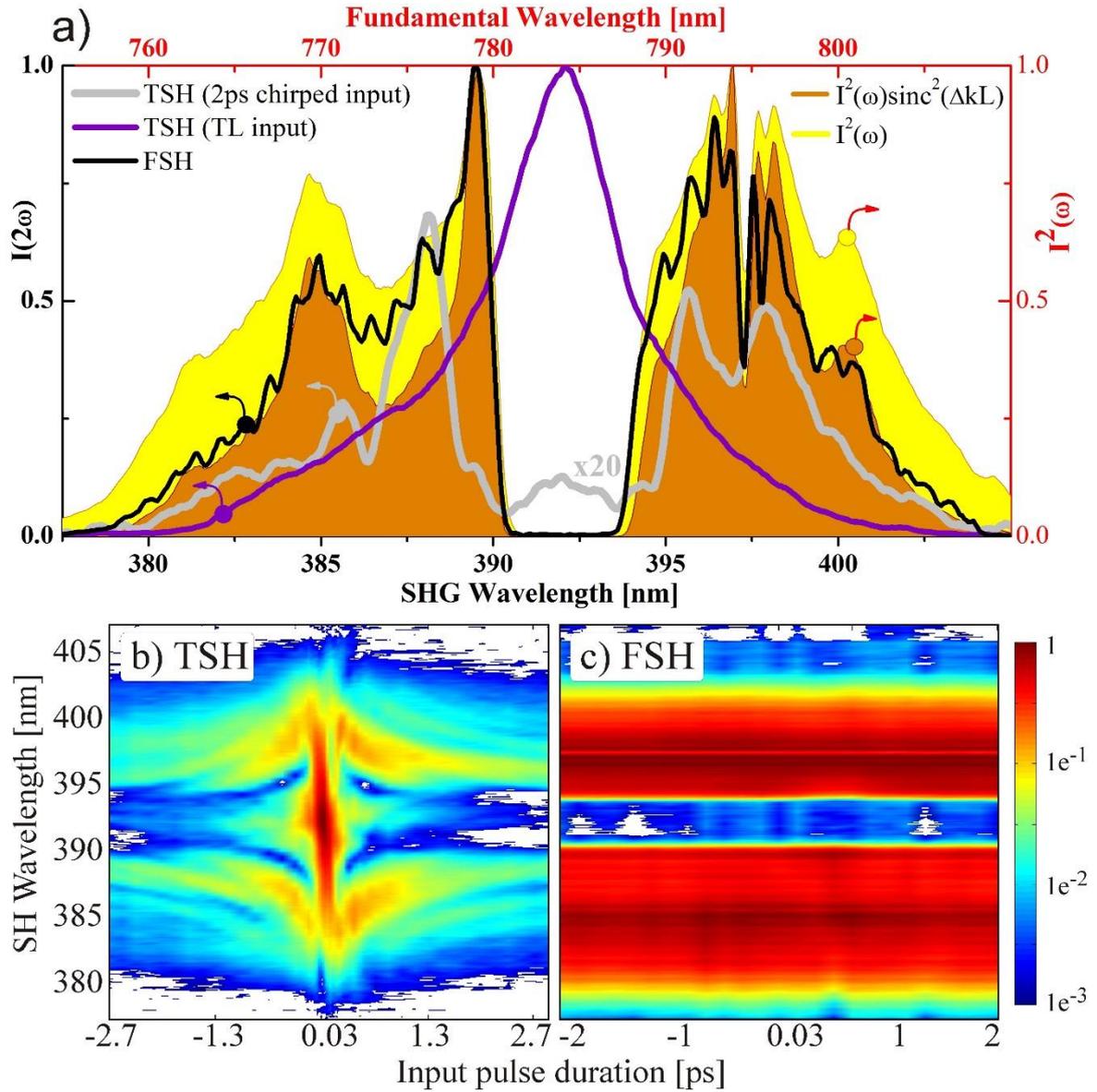

Figure 4: Investigating amplitude and phase coupling as a function of chirp of the fundamental pulse. The fundamental amplitude was shaped according to the yellow curve in a) showing its intensity squared. TSH (b) exhibits a strong dependence of the spectral amplitude as a function of input chirp. Because this amplitude-phase coupling is absent in FSH (c), all SH spectra remain constant, like the black curve in a). All spectral features transfer directly to the SH. Since it is difficult to analyse feature sharpness in TSH where the SH spectrum strongly depends on the input shape and phase, we investigate the time span required to develop the spectral hole introduced in the input beam. The TL duration of the fundamental is about 33 fs FWHM. The missing central part corresponds to a TL about 360 fs FWHM (zero-zero duration ~1 ps). The grey curve in (a) for a chirped input of ~2 ps still shows features at the central wavelength indicating that long temporal stretching is required to minimize the mutual coupling of frequencies in TNO. Thus, efficient amplitude and phase coupling can only be turned off in FSH.

*Future prospects*

To battle the loss of bandwidth due to phase mismatch, which seems even more pronounced in TSH than FSH (compare black curve to grey and violet ones in Fig. 4 (a)), TSH offers the choices of using thinner crystals with less efficiency or elaborate achromatic phase matching schemes[11]. While a variation of the later can also be realized in FSH it also offers new implementations such as specially periodically poled materials[12] or the use of the separation Ansatz[13] with multiple crystals to efficiently double an octave of frequencies. While the time domain frequency conversion of chirped pulses like for the grey curve of Fig. 4(a) may be improved following a more complex adiabatic scheme[14], the problem of recompression afterwards still remains. Another application of our scheme could be the coherent combination of single-cycle fundamental with single-cycle SH pulses to generate multi-octave sub-cycle light fields[15] or to widen frequency bands for metrology[16]. On the other hand, the arbitrary phase transfer would allow straightforward pulse shaping all the way down to 200 nm[17] with a conventional shaper in the Ti:Sa fundamental beam. Linearized nonlinear optics could extend the functionality of integrated optical devices[18] and logical operations[19,20].

We also envision a modified type of 2D spectroscopy[3] *by* placing the sample in the FP and utilizing the spread colors as a narrow band pump in each focal spot. Next, all spots can be simultaneously probed by a time delayed fs pulse. The FP can be re-imaged onto another FP where all emission spectra for each excitation spectrum of the first FP can be detected in a single shot.

Ultimately, a large number of nonlinear effects discovered during the last half century can be revisited in the frequency domain, to generate different results based on the elimination of nonlinear frequency couplings in nonlinear interactions and the possibility of direct phase and amplitude transfer to other wavelengths.

**Acknowledgements** We thank Misha Ivanov, Benjamin A. Schmidt and David M. Villeneuve for valuable discussions. We acknowledge financial support from the NSERC, FRQNT, MEIE and CFI.

**Author Contributions** B.E.S., M. C., H. I. and F. L. conceived the experiments; P. L., G. E., M. C. and B.E.S. performed the experiments; B.E.S., M. C. and G. E. developed the theory; allauthors contributed to the manuscript.

Methods

**Laser system**
The measurements were carried out with 1 mJ, 33 fs pulses at 780 nm wavelength from a Ti:Sa amplifier operating at 500 Hz at the advanced Laser Light Source (ALLS). The beam diameter was 4 mm FWHM of intensity carrying 500 µJ of pulse energy. The amplitude- and phase-shaping was carried out with an acousto-optic programmable dispersive filter (AOPDF). The results of figure 4 showing the chirp dependence were performed by changing the grating separation in the compressor since it was difficult to obtain huge amounts of chirp with the AOPDF. Spectral measurements were carried out with an integration sphere fiber-coupled to an imaging spectrometer. Pulse characterization was performed with a special, all reflective transient grating – frequency resolved optical gating (TG-FROG)[10] apparatus. Beam separation of the three beams was carried out with a transmission mask consisting of three holes which enabled wavelength and polarization independent operation[21]. The same device can measure the fundamental and SH pulses regardless of their polarization. Unlike SHG-FROG, it has no time ambiguity and generates intuitive spectrograms for quadratic and cubic phase functions which facilitates phase retrieval and data interpretation.

*4f* **setup**

A *4f* setup comprising reflection gratings (600 lines/mm) and lenses (f=300mm) as shown in figure 1c) was used in the experiment. Additionally a half wave plate was inserted prior to L1 to rotate the initially perpendicular polarization of the fundamental by 90°. This sole modification is not shown in figure 1(c) for clarity. Since a 150µm thick type I BBO is used for doubling, the SH polarization is perpendicular to the fundamental. The half wave plate between the gratings ensures perpendicular polarization of the SH beam on the exit grating. This allows angle tuning of BBO phase matching along the spectrally dispersed axis. Additionally, one can achieve more efficient grating operation at the SH wavelength if the polarization is perpendicular to the grating grooves. The long focal lengths of L1 & L2 lead to a rather low intensity in the FP (around 1GW/cm$^2$). This, in combination with the thin BBO (150µm) yielded an SHG efficiency in the FP of about 5%. The setup was not optimized for SHG efficiency (low intensity in the FP) and the thin BBO was chosen because thicker crystals showed pronounced phase matching deficiencies that make comparison between FSH and TSH difficult.

The long focal lengths also eased modifications and allowed access to the FP. The efficiency should increase with shorter focal lengths, higher pulse energy or thicker crystals. The overall SH efficiency might be further improved by choosing a grating blazed for 400nm with twice the groove size of the input grating satisfying: $\sin(\alpha)\lambda_{800nm}/g = \sin(\alpha)\lambda_{400nm}/2g$. In this way, the fundamental and the SH beam would be separated by the grating. However, we chose the same grating at the exit side for convenience and because it can simultaneously recombine the fundamental and SH beam. The SH beam recombines through 2$^{nd}$ order diffraction while the fundamental beam recombines via 1$^{st}$ order. The SH beam exhibits no spatial chirp and the far field measurement shows a perfectly symmetric focal spot containing all spectral components (see SI). We found, however, that fundamental and SH beams recombine optimally for slightly different incident angles on the second grating G2. We note, that this might be related to the angular acceptance bandwidth of BBO. The bandwidth and corresponding pulse duration in the focal spot of the FP depend on the geometric optical properties of the setup[13]:

$$\Delta\lambda_{foc} = \frac{\lambda_c\,g}{\pi\,d_{in}} 2\ln(2) \sqrt{1 - \left(\frac{\lambda_c}{2g}\right)^2}$$

$$\Delta t_{foc} = \frac{\lambda_c\,d_{in}}{g\,c} \left[1 - \left(\frac{\lambda_c}{2g}\right)^2\right]^{-1/2}$$

In typical experimental arrangements, the focal bandwidth in the FP is > 100 times smaller than the input pulse spectrum. The finest spectral feature that can be transferred via FSH is given by the spectral resolution $\Delta\lambda_{foc}$ of the *4f* setup.

# Supplementary information "Linearizing Nonlinear Optics"

**SI I) Nonlinear interaction in the Frequency domain.**

This section is dedicated to the derivation of Eq. (4) describing the nonlinear interaction in the Frequency domain. In the Fourier plane (FP), the initial spectrum is spatially separated such that only a "single" frequency $w'$ is present in each focal spot. Therefore the spectrum in one focal spot can be expressed as the product of the initial spectrum and a delta distribution: $E(w) \cdot \delta(w - w')$. This reduced bandwidth interacts nonlinearly in the BBO crystal. The nonlinear interaction is represented by a convolution in the frequency representation like in Eq. (2). Finally, the last grating recombines all frequency together which is represented by an integral over all $w'$. Thus, the nonlinear interaction in a *4f* setup can be expressed as:

$$E_F^{SH}(w) \sim \int_{-\infty}^{\infty} dw' \, \bigl(E(w) \cdot \delta(w - w')\bigr) * \bigl(E(w) \cdot \delta(w - w')\bigr). \tag{SI 1}$$

After performing the multiplication with the Dirac delta, only the convolution of two deltas remains. Note that $E(w')$ is not a function of $w$ any more:

$$E_F^{SH}(w) \sim \int_{-\infty}^{\infty} dw' \, E(w') \cdot E(w') \cdot \delta(w - w') * \delta(w - w'). \tag{SI 2}$$

According to the convolution properties of the Dirac delta $\delta(x - a) * \delta(x - b) = \delta(x - a - b)$, this equation can be written as follows:

$$E_F^{SH}(w) \sim \int_{-\infty}^{\infty} dw' \, E(w') \cdot E(w') \cdot \delta(w - 2w'). \tag{SI 3}$$

$$E_F^{SH}(w) \sim \int_{-\infty}^{\infty} dw' \, E(w') \cdot E(w') \cdot \delta\left(-2\left(w' - \frac{w}{2}\right)\right). \tag{SI 4}$$

With $\delta\left(\frac{x}{b}\right) = |b|\delta(x)$

$$E_F^{SH}(w) \sim \int_{-\infty}^{\infty} dw' \, E(w') \cdot E(w') \cdot \delta\left(w' - \frac{w}{2}\right). \tag{SI 5}$$

$$E_F^{SH}(w) \sim \int_{-\infty}^{\infty} dw' \, E\left(\frac{w}{2}\right) \cdot E\left(\frac{w}{2}\right) \cdot \delta\left(w' - \frac{w}{2}\right). \tag{SI 6}$$

This regular convolution integral is evaluated using the properties of the Dirac delta and leads to:

$$E_F^{SH}(w) \sim E\left(\frac{w}{2}\right) E\left(\frac{w}{2}\right) \tag{SI 7}$$

Therefore, the spectrum in time domain second harmonic is the square of the initial spectrum. The *w/2* term (represents the fundamental frequency) accommodates the fact that the frequency axis needs to be expand by a factor of 2. This will have the effect to broaden the function by a factor of 2 and to move the center of the function from the fundamental to its second harmonic which is an expected result of the SHG. The broadening by a factor of 2 holds for $E_F^{SH}(w)$ and $I_F^{SH}(w)$ will be broader by $\sqrt{2}$ as expected for a SH process in the absence of phase matching restrictions. In Eq. (4) of the main paper we substituted $\omega = w/2$.

## SI II) Approximation using δ function

One immediate concern to be raised is the validity of the "single" frequency assumption in the FP on which the introduction of the Kronecker delta function is based. We will discuss two approaches to justify this assumption.

First, from a geometrical optical point of view. The 4f setup not only performs a temporal Fourier transformation from time to frequency domain but also a spatial Fourier transformation. The latter merely denotes the main purpose of a focusing optic, it focuses a parallel beam into a focal spot or collimates a point source into a collimated beam, respectively. This point source can be seen as a delta function whose Fourier transform yields a constant function, the plane wave. This idealized picture disregards the effect of finite optic sizes. If the 4f setup is illuminated with a polychromatic plane wave, each frequency leaves the grating under a different diffraction angle while still being a plane wave. The focusing optic performs a spatial Fourier transformation of each plane wave to generate a delta function. An inclined, plane wave incident onto a lens (i.e. different pulse frequencies) corresponds to a function with a linear phase slope with respect to the spatial coordinate. This linear phase slope causes a displacement in the Fourier transform plane, thus all frequencies get spatially separated.

Secondly, to demonstrate the validity of the delta function approximation, we perform numerical simulations according to Eq. (SI 1) assuming a variable focal spot size in the FP by replacing the delta function by a Gaussian with variable width. In other words, we now take into account the influence of finite optic sizes. Changing the focal spot size is equal to a change of the effective spectral bandwidth for the nonlinear interaction. The FWHM bandwidth of a TL 33fs pulse corresponds to 28nm. In the experiment, the FWHM bandwidth in each focal of the FP spot was roughly 0.1nm, i.e. about 300 times smaller than the pulse bandwidth. Table SI 1 shows the evolution from FNO to TNO as the effective bandwidth in the 4f setup approaches FWHM bandwidth of the pulse.

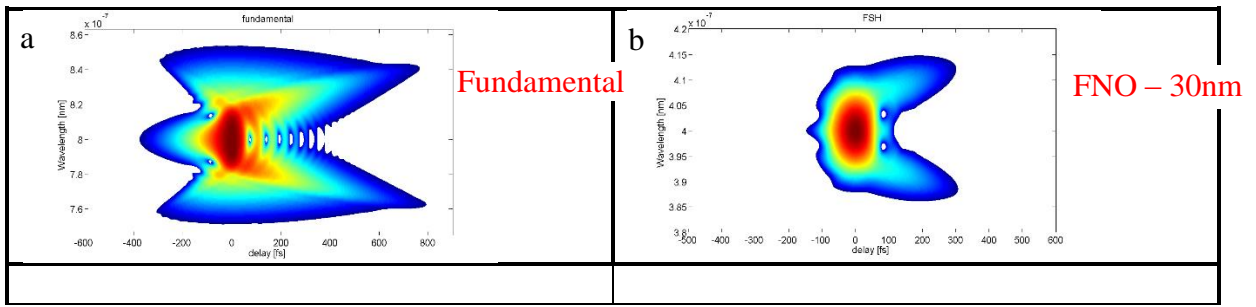

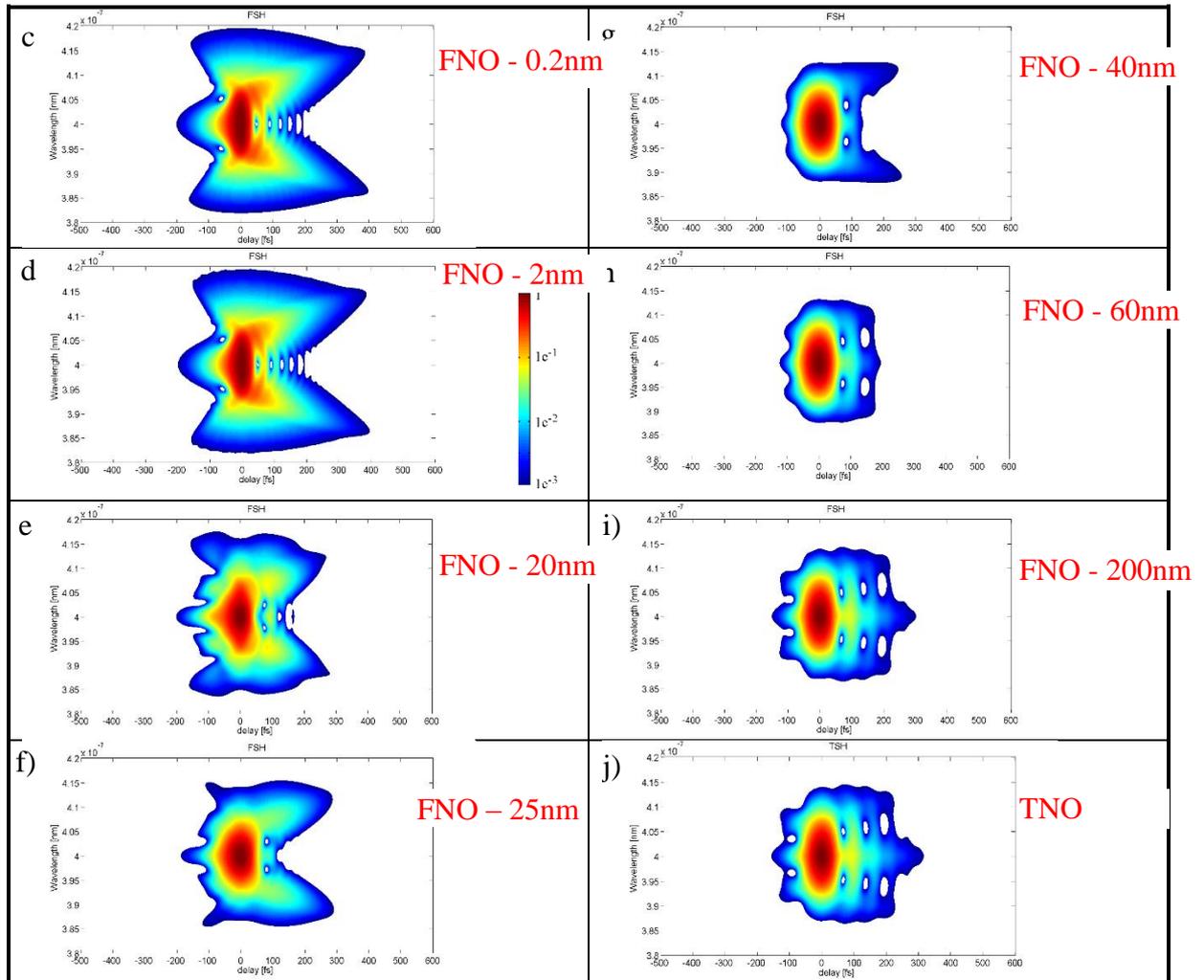

Table SI 1: Spectrograms showing the influence of varying resolution of the 4f setup in the Fourier plane (b-i). (a) shows the fundamental pulse with a FWHM of 28nm and (j) the time domain SHG. The experimental resolution of the 4f setup was 0.1nm. As this resolution is increased, one can follow the transition from FNO to TNO. Note the logarithmic scale.

For a low enough resolution of the 4f setup, the result of FNO equals TNO (compare i) and j)). This means, Eq. (3) of the main paper is a more general description for nonlinear optical interaction with time domain interaction denoting a limit case of FNO with largely reduced spectral resolution.

**SI III) Transfer of other phase functions**

As mentioned in the main paper, (only) second order phases can be transferred via TNO which is confirmed by the experimental spectrograms in Fig. SI 1 (b) & (d). The different sign of the fundamental input chirp (not shown) is transferred to the SH. The same is true for FNO.

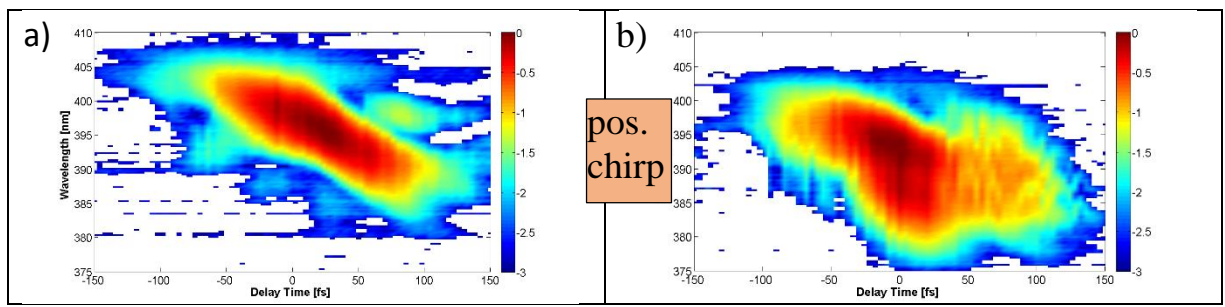

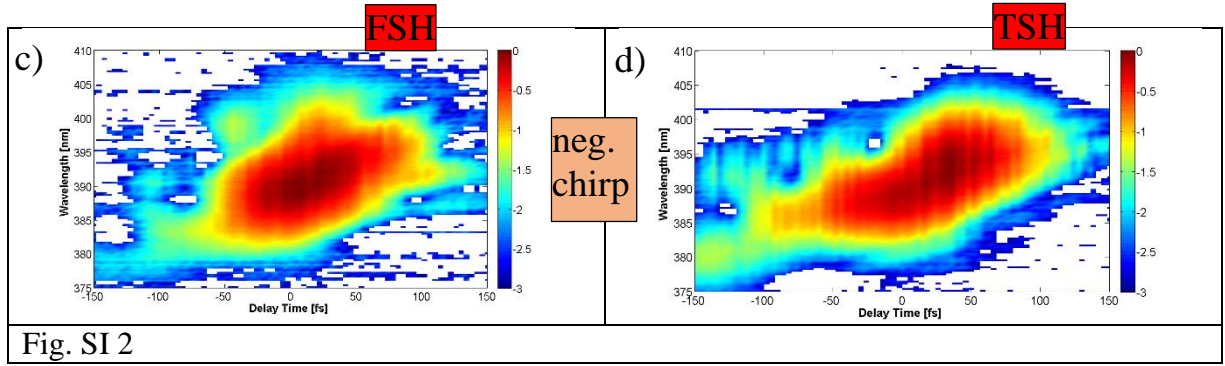

Fig. SI 2

For completeness, we also show the spectrograms for opposite TOD sign as compared to the results of Fig. 2 in the main paper. Due to experimental deficiencies, the quality of phase shaping is reduced. The reason is that we aligned the acoustic wave settings inside the AOPDF such that it enables maximum amount of chirp in one direction (+ 150 000 fs3).

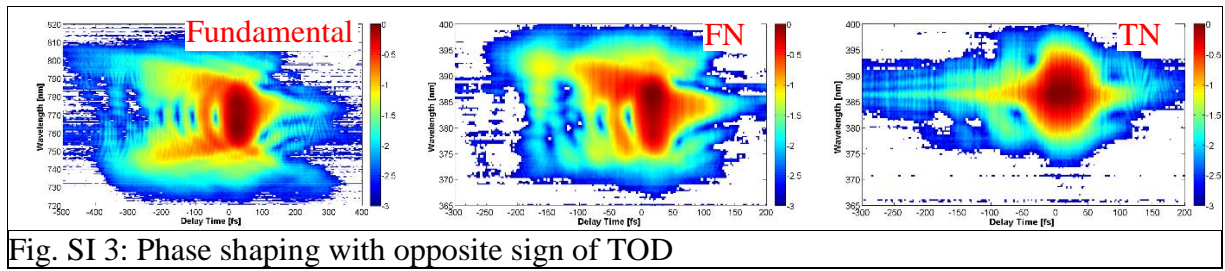

Fig. SI 3: Phase shaping with opposite sign of TOD

## SI IV) Role of phase matching

In a macroscopic medium, there are two limiting factors for the bandwidth besides absorption. First, on the microscopic scale it is the photon order of the process involved. A SH process requires two fundamental photons for the emission of a photon at twice the photon energy. This quadratic dependence of output power spectral density on input intensity leads to a spectral narrowing which is illustrated in Fig SI 3. The blue curve shows the square of the input spectrum in red. The amount of narrowing depends on the shape of the function involved.

Secondly, in the case of an extended medium, macroscopic propagation effects will occur. The most important one is phase matching which is required for a constructive superposition of coherent waves. The maximum power spectrum that is phases matched in SHG based on birefringent phase matching is given by[1] $sinc^2(\Delta k L)$, plotted as the green curve in Fig. SI 3. L is half the crystal thickness and $\Delta k$ is the wave vector mismatch between all three interacting waves.

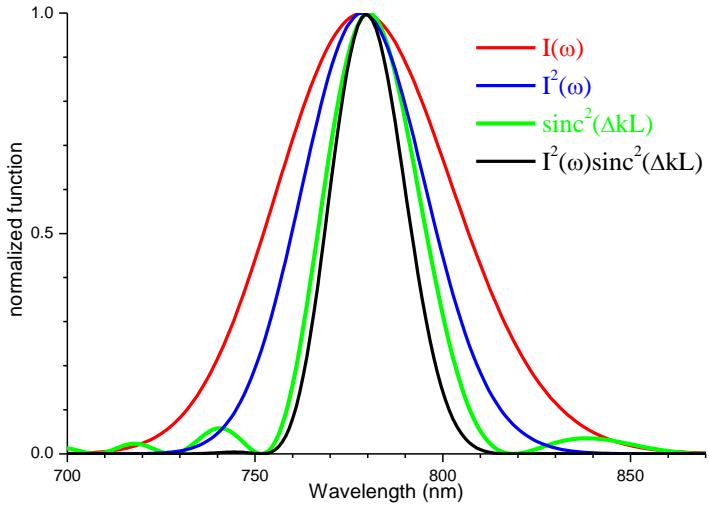

Fig. SI 4: Incident bandwidth (red) and phase matched bandwidth (black)

The final phase matched bandwidth for a SH process is the multiplication of both effects: $I^2(w)sinc^2(\Delta kL)$, plotted as the black curve. We point out that this restriction applies to both TNO and FNO. FNO, however, offers the chance to double a higher relative bandwidth than TNO if a spectral shape different than Gaussian is chosen as the input function.

### SI V) 4f setup output properties

As mentioned in the main paper, the output properties of the 4f setup at the SH wavelength are very witnessed by the absence of spatial chirp and a good focusing properties. An image of the focal spot is shown in Fig. SI 5.

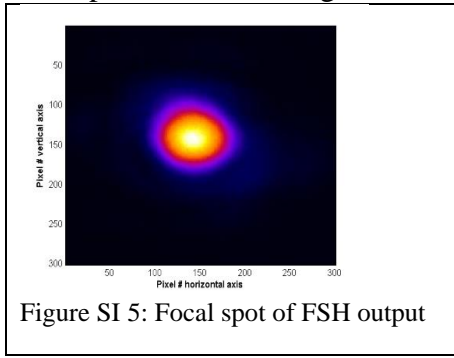

Figure SI 5: Focal spot of FSH output

Another indication of good spatial quality are the TL limited pulse duration measured with the TG-FROG which uses a plat with three spatially separated holes to provide the three beams for the four wave mixing process[2]. Only if the spectral contend across the entire beam is spatially uniform, it is possible to measure TL pulse durations. The otherwise reduced bandwidth leads to longer pulse durations.